# Experimental Characterisation of Distributed Reactive Power Sharing under Communication-Induced Stress in Parallel Grid-Forming Inverters

Edgar Diego Gomez Anccas[1*], Eilean MacPherson[1], Jirko Tegeler[1],
Kevin Röbert[2], Mathias Fischer[2], Christian A. Hans[3], Detlef Schulz[1]
[1]Chair of Electrical Power Systems, Helmut Schmidt University, Hamburg, Germany
[2]Computer Networks, University of Hamburg, Germany
[3]Automation and Sensorics in Networked Systems Group, University of Kassel, Germany
[*]Corresponding author: diego.gomez@hsu-hh.de

*Abstract*—Synchronisation of parallel grid-forming inverters is crucial for stable operation of future power systems. This includes accurate and robust reactive power sharing under realistic operating conditions such as impedance mismatch and communication constraints. In this work, reactive power sharing by virtue of a distributed control law is investigated under line impedance mismatch. Furthermore, robustness and transient behaviour of the proposed approach are experimentally evaluated under communication-induced stressors including a fixed 3 % packet loss and communication delays ranging from 50 ms to 100 ms, artificially introduced through a software-defined overlay. The study is conducted in a low-voltage laboratory-scale microgrid comprising two parallel grid-forming inverters, an AC load, and a grid-following battery system acting as a reactive power injector. The results show reactive power sharing convergence up to 90 ms communication delay, with a stability boundary between 90 ms and 100 ms, which decreases with increasing integral gain.

*Index Terms*—Grid-forming inverters, reactive power sharing, microgrid, communication delay, distributed secondary control.

## I. INTRODUCTION

Increasing inverter penetration as a consequence of the integration of renewables in modern power systems introduces several challenges. In particular, future power systems are subject to increased uncertainty and require sophisticated control strategies since power electronic converters lack the inherent physical advantages of rotating machines such as mechanical inertia and self-synchronisation. Different inverter control approaches have been investigated such as grid-following inverter (GFLI) and grid-forming inverter (GFMI) control [1], [2]. The key distinction between these approaches lies in how they interact with the electrical system. GFLIs require an existing grid to operate. The synchronisation is realised using techniques based on phase-locked loops (PLLs) to adjust output voltage magnitude and phase to the grid. In contrast, GFMIs can operate in standalone conditions or weak grids by establishing their own voltage and frequency. The distinction between GFMIs and GFLIs is well established and several solutions exist for the synchronisation of GFLIs with existing grids. Existing work focuses on switching between both operating modes [3], on the synchronisation of GFLIs [4] and on stability of GFMIs and GFLIs in parallel operation [5]. One study that explicitly addresses the synchronisation of parallel GFMIs is presented in [6]. In this work, two distinct control approaches for synchronising two GFMIs are proposed, i.e., output sync and control sync, and validated experimentally using a laboratory setup. While the study focuses on stability, accurate reactive power sharing among parallel GFMIs is not investigated.

Conventional droop-based strategies, exhibit sensitivity towards line impedance mismatch, which can result in steady-state reactive power deviations. Recent publications have proposed several methods to improve reactive power sharing accuracy in standalone AC microgrids. Sliding mode-based strategies have demonstrated enhanced robustness against parameter variations and impedance mismatch [7]. Moreover, adaptive virtual impedance techniques, which modify the effective output impedance to improve the precision of power sharing, have been shown to be useful [8], [9]. Furthermore, advanced distributed secondary control strategies to enhance voltage and frequency regulation as well as power sharing accuracy have been introduced. These include Lyapunov function-based control strategies [10], fixed-time control methods [11], delay-compensated controller designs based on bi-limit homogeneity theory [12] and hierarchical robust architectures that integrate model-free predictions with fixed-time control to improve dynamic performance [13]. Other publications focus on communication-related challenges. Resilient distributed secondary control strategies have been proposed in [14] to ensure coordinated operation under malicious data injection attacks. Moreover, diffusion-based distributed secondary control methods have been compared to consensus algorithms, demonstrating improved convergence rates and robustness under simu-

This research paper is funded by dtec.bw–Digitalization and Technology Research Center of the Bundeswehr–which we gratefully acknowledge. dtec.bw is funded by the European Union–NextGenerationEU.

lated communication delays and link failures in [15]. Although the aforementioned publications provide valuable contributions in the field of accurate reactive power sharing, distributed secondary control design and communication robustness, the majority emphasises controller synthesis or simulation-based validation. However, systematic experimental characterisation of stability limits based on communication-induced stressors using realistic communication structures remains an open topic.

We aim to fill this gap by experimentally investigating distributed reactive power sharing between GFMIs in parallel operation under stressed communication conditions such as increasing time delays, packet loss and fixed update rates, which we artificially alter in a software-defined overlay (SDO). A distributed integral correction scheme is implemented and evaluated on real laboratory-scale microgrid consisting of two 15 kW-rated GFMIs with DC/DC and DC/AC stages and a 400 V battery-based GFLI system. The interaction between integral gain selection and communication delay is systematically analysed to identify gain- and delay-dependent stability bounds and provide practical insight into distributed reactive power sharing under communication constraints.

The remainder of this paper is structured as follows. In Section II, the control scheme for the parallel operation of GFMIs is introduced. Section III presents the distributed reactive power sharing controller and discusses the associated communication requirements and the implemented SDO. In Section IV, the experimental low-voltage microgrid case study is described and the experimental results are presented. Finally, Section V concludes the paper.

## II. Control of Parallel Grid-Forming Inverters

The fundamental cascaded control structure for parallel operation consists of a droop control stage followed by inner voltage and current control loops (see Fig. 1). In what follows, the control structure and the synchronisation method are presented.

### A. Droop-based Grid-Forming Control

For grid-forming operation, a sequence of control actions is applied to generate adequate duty cycles $D \in \mathbb{R}^3$ for the inverter. First, droop control is implemented to generate voltage and frequency references for the system via

$$\omega = \omega^* - k_{\mathrm{drp}}(P - P^*), \tag{1a}$$
$$v_{\mathrm{d}}^* = v_{\mathrm{m}}^* - k_{\mathrm{drq}}(Q - Q^*), \tag{1b}$$
$$v_{\mathrm{q}}^* = 0, \tag{1c}$$
$$\dot{\theta} = \omega. \tag{1d}$$

Here, $\omega$ denotes the angular frequency, $v_{\mathrm{d}}^*$ and $v_{\mathrm{q}}^*$ are the voltage references in the $dq$ reference frame, $v_{\mathrm{m}}^*$ and $\omega^*$ are the nominal voltage magnitude and angular frequency references, respectively. Moreover, $P$ and $Q$ are the active and reactive power measurements, respectively. Their corresponding reference values are $P^*$ and $Q^*$. The $dq$ reference frame is aligned with the inverter voltage vector such that (1c) holds. Furthermore, balanced load conditions are assumed during operation.

### B. Voltage and Current Control Cascades

For good control performance, voltage and current control loops are set up according to [16]. The input of the outer voltage controller is the computed voltage reference $v^* = \begin{bmatrix} v_{\mathrm{d}}^* & v_{\mathrm{q}}^* \end{bmatrix}^\top$, which is used to generate the current reference $i_{\mathrm{s}}^*$ via

$$i_{\mathrm{s}}^* = \begin{bmatrix} k_{\mathrm{pvd}}(\Delta v_{\mathrm{d}}) + k_{\mathrm{ivd}} \int_0^t \Delta v_{\mathrm{d}} \mathrm{d}t + i_{\mathrm{d}} - v_{\mathrm{q}} C_{\mathrm{g}} \omega \\ k_{\mathrm{pvq}}(\Delta v_{\mathrm{q}}) + k_{\mathrm{ivq}} \int_0^t \Delta v_{\mathrm{q}} \mathrm{d}t + i_{\mathrm{q}} + v_{\mathrm{d}} C_{\mathrm{g}} \omega \end{bmatrix}. \tag{2}$$

Here $\Delta v = \begin{bmatrix} \Delta v_{\mathrm{d}} & \Delta v_{\mathrm{q}} \end{bmatrix}^\top = v^* - v$ is the error between voltage reference and the measurement values $v = \begin{bmatrix} v_{\mathrm{d}} & v_{\mathrm{q}} \end{bmatrix}^\top$ in the $dq$ reference frame. Moreover, $i = \begin{bmatrix} i_{\mathrm{d}} & i_{\mathrm{q}} \end{bmatrix}^\top$ is the current measurement in the $dq$ frame, $C_{\mathrm{g}}$ is the output filter capacitance and $k_{\mathrm{pvd}}$, $k_{\mathrm{ivd}}$, $k_{\mathrm{pvq}}$ and $k_{\mathrm{ivq}}$ are the PI gains for the $d$- and $q$-axes. Using the current reference, the switching voltage reference $v_{\mathrm{s}}^*$ is then computed via

$$v_{\mathrm{s}}^* = \begin{bmatrix} k_{\mathrm{pid}}(\Delta i_{\mathrm{sd}}) + k_{\mathrm{iid}} \int_0^t \Delta i_{\mathrm{sd}} \mathrm{d}t + v_{\mathrm{d}} - i_{\mathrm{sq}} L_{\mathrm{g}} \omega \\ k_{\mathrm{piq}}(\Delta i_{\mathrm{sq}}) + k_{\mathrm{iiq}} \int_0^t \Delta i_{\mathrm{sq}} \mathrm{d}t + v_{\mathrm{q}} + i_{\mathrm{sd}} L_{\mathrm{g}} \omega \end{bmatrix}, \tag{3}$$

where $\Delta i_{\mathrm{s}} = \begin{bmatrix} \Delta i_{\mathrm{sd}} & \Delta i_{\mathrm{sq}} \end{bmatrix}^\top = i_{\mathrm{s}}^* - i_{\mathrm{s}}$ is the error between the current reference and their respective filter current measurement $i_{\mathrm{s}} = \begin{bmatrix} i_{\mathrm{sd}} & i_{\mathrm{sq}} \end{bmatrix}^\top$ in the $dq$ reference frame. Moreover, $L_{\mathrm{g}}$ is the output filter inductance and $k_{\mathrm{pid}}$, $k_{\mathrm{iid}}$, $k_{\mathrm{piq}}$ and $k_{\mathrm{iiq}}$ are the PI gains for the $d$- and $q$- axes. Lastly, the duty cycle is computed by transforming $v_{\mathrm{s}}^*$ to the abc reference frame with the inverse park transformation using $\theta$ yielding $v_{\mathrm{sabc}}^*$ and then normalising it by the DC link voltage $v_{\mathrm{dc}}$, i.e.,

$$D = \frac{v_{\mathrm{sabc}}^*}{v_{\mathrm{dc}}}. \tag{4}$$

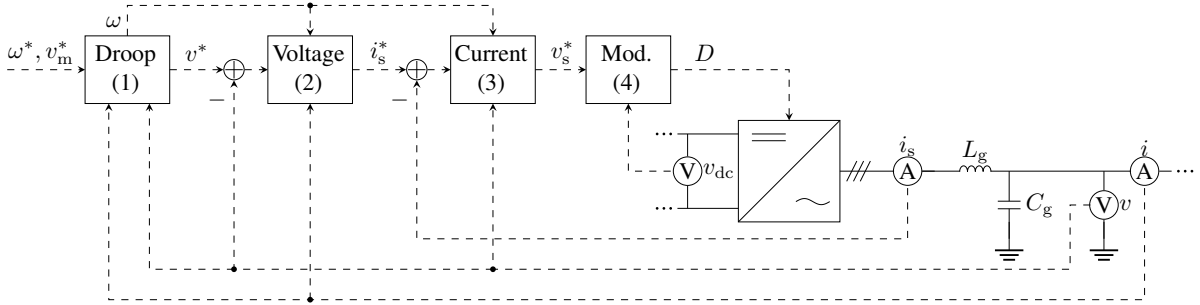

Fig. 1. Cascaded inverter control composed of droop control (1), voltage control (2), current control (3) and duty cycle computation (4).

Implementing this can reliably setup a grid. However, for the parallel operation of multiple GFMIs, further synchronisation is required. The system architecture considered in this study is shown in Fig. 2.

### C. Synchronisation of Parallel Grid-Forming Inverters

For a successful connection of GFMIs to an existing grid, the connecting unit has to first match the voltage magnitude and phase angle and then, after connection participate in active and reactive power sharing. To align voltage magnitude and phase angle of the $n$-th inverter, based on [6], (1a) and (1b) are extended by additional correction terms for frequency $\delta_{\omega n}$ and voltage magnitude $\delta_{vn}$, i.e.,

$$\omega_n(t) = \omega^* - k_{\mathrm{drp}}(P_n(t) - P_n^*) - \delta_{\omega n}(t), \tag{5a}$$
$$v_{\mathrm{dn}}^*(t) = v_{\mathrm{m}}^* - k_{\mathrm{drq}}(Q_n(t) - Q_n^*) - \delta_{vn}(t). \tag{5b}$$

The connecting GFMI requires the voltage measurements of the established AC system $v_g$ to compute the correction terms. The voltage is transformed into $dq$-frame using the phase $\theta_n$ of the connecting GFMI creating the shifted system $v_{\mathrm{dq}}^{(n)}$. The shifted system's $q$ component is adjusted to zero by regulating $\delta_{\omega n}$ using a PI controller

$$\delta_{\omega n}(t) = k_{\mathrm{p}\omega} e_{\omega n}(t) + k_{\mathrm{i}\omega} \int_0^t e_{\omega n}(\tau) \mathrm{d}\tau, \tag{6}$$

to match the phases as proposed in [6]. Here, $k_{\mathrm{p}\omega}$ and $k_{\mathrm{i}\omega}$ denote the PI gains and $e_{\omega n}$ the state-dependent phase error term,

$$e_{\omega n}(t) = \begin{cases} v_{\mathrm{q}}^{(n)}, & \text{if } S_n = 0, \\ -k_{\mathrm{fb}}\,\delta_{\omega n}(t), & \text{else}, \end{cases} \tag{7}$$

with feedback gain $k_{\mathrm{fb}}$. During synchronisation, the relay $S_n = 0$ is open to align the phases. Once synchronisation is complete, $S_n = 1$ is closed, thereby gradually eliminating the correction term and enabling power sharing.

Similarly, the voltage magnitude is matched via

$$\delta_{vn}(t) = k_{\mathrm{pv}} e_{vn}(t) + k_{\mathrm{iv}} \int_0^t e_{vn}(\tau) \mathrm{d}\tau, \tag{8}$$

with PI gains $k_{\mathrm{pv}}$ and $k_{\mathrm{iv}}$ and the state dependent voltage magnitude error term

$$e_{vn}(t) = \begin{cases} v_{\mathrm{fil}}(t) - v_{\mathrm{d}}^*(t), & \text{if } S_n = 0, \\ -k_{\mathrm{fb}}\delta_{vn}(t), & \text{else}. \end{cases} \tag{9}$$

Here $v_{\mathrm{fil}}$ denotes the filtered voltage magnitude of the established AC system.

## III. REACTIVE POWER SHARING CONTROLLER

This section introduces the distributed reactive power sharing controller used to compensate sharing errors caused by line impedance mismatch. Furthermore, the associated communication requirements and the implemented solution are presented.

### A. Reactive Power Sharing Control Law

While the synchronization and active power sharing can be achieved through the aforementioned approach, reactive power sharing is not guaranteed in presence of impedance mismatch. Therefore, motivated by [17], the droop relations in (5b) are extended by a distributed reactive power sharing correction term $\delta_{Qn}$. After synchronisation, the correction terms $\delta_{\omega n}$ and $\delta_{vn}$ converge to zero. Consequently, both units follow the new voltage control law, which for a system of $N$ parallel inverter and $\forall n \in \{1, \ldots, N\}$ reads

$$v_{\mathrm{d}n}^*(t) = v_{\mathrm{m}}^* - k_{\mathrm{dq}}(Q_n(t) - Q_n^*) - \delta_{vn}(t) - \delta_{Qn}(t), \quad (10\mathrm{a})$$

$$\delta_{Qn}(t) = k_{\mathrm{iQ}} \int_0^t (\overline{Q}_{\mathrm{sn}}(t) - Q_n(t))\mathrm{d}t, \quad (10\mathrm{b})$$

$$\overline{Q}_{\mathrm{sn}}(t) = Q_n^* + \rho_n \sum_{i=1}^N (Q_i(t) - Q_i^*) \quad (10\mathrm{c})$$

$$\sum_{n=1}^N \rho_n = 1, \quad (10\mathrm{d})$$

with $k_{\mathrm{iQ}}$ as the integral gain, $\overline{Q}_{\mathrm{sn}}$ as the computed reactive power sharing target and $\rho_n$ as the contribution factor of the $n$-th inverter. The integral term ensures convergence of each inverters' reactive power output $Q_n$ towards their weighted contribution. In contrast to pure droop control, steady-state sharing errors caused by system asymmetries can be eliminated. However, as the correction term relies on the exchange of locally computed reactive power values between inverters, the closed-loop performance becomes sensitive to communication delay and update rate.

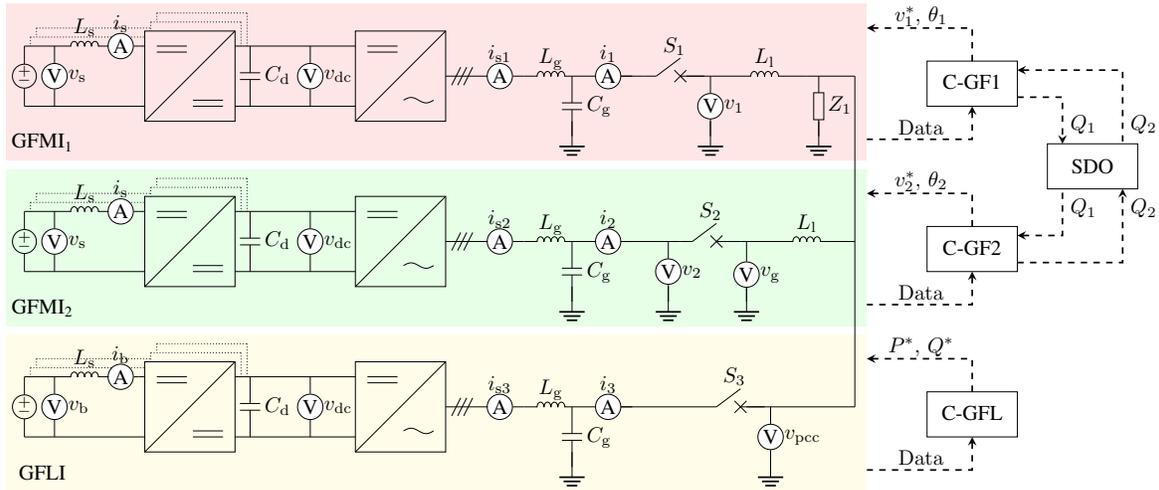

Fig. 2. System architecture comprising two parallel GFMIs (GFMI$_1$, GFMI$_2$) and one GFLI with local primary controllers and an SDO link between the GFMI controllers.

### B. Communication Requirements

Although (10) displays a straightforward way to address reactive power imbalance, it relies heavily on coordination and communication between the individual units. As the timescale of the communication is significantly slower than the voltage and current control loops, communication via an SDO [18] channel is employed to ensure authenticity and integrity of the communication channel. An SDO further enables the integration of network modules tailored to our requirements, namely a reliable, secure, and authentic communication channel with minimal delay. Within an SDO, we can specify the properties of an ideal network, such as the absence of firewalls, direct communication paths to minimise latency, and encrypted and authenticated communication. The overlay software then translates these requirements into the necessary actions on the underlying physical network topology, as an example creating direct peer-to-peer connections via NAT-hole punching techniques. In contrast, traditional communication paradigms require communication through a publicly reachable relay in the presence of firewalls, which increases packet round trip time thus introducing a delay into our control approach. SDOs have demonstrated their ability to minimize latency compared to traditional networking [19]. Additionally, an SDO allows arbitrary manipulations of the (virtual) network link, such as increasing delay and packet loss to simulate network degradations.

## IV. CASE STUDY

This section presents the experimental case study used to characterise the system's behaviour under the proposed control strategy and communication stressors. First, the experimental setup is described. Next, five experiments are conducted to investigate the baseline droop behaviour, the effect of impedance mismatch, the performance of the reactive power correction term, the sensitivity of the integral gain of said term and the robustness against communication-induced stressors.

### A. Experimental Setup

The experimental setup extends the configuration presented in [20]. It consists of two identical 15 kW-rated grid forming units GFMI$_1$ and GFMI$_2$ connected in parallel as depicted in Fig. 3. Each unit comprises three DC sources connected to a common DC bus through an interleaved boost converter, increasing the DC bus voltage to 700 V. Details of the DC bus control are provided in [21]. Moreover, the reactive power sharing extension (10) is implemented on both controllers and can be activated or deactivated externally.

The GFLI string shares the same power-stage architecture but uses an 11.6 kWh, 400 V lithium-ion battery system as DC source. The controller of each string is programmed using ACG-SDK 2025.1.2 in conjunction with Simulink 2024b and is executed on an imperix B-Box RCP 3.0 operating at a switching frequency of $f_\mathrm{s} = 20\,\mathrm{kHz}$ [22], [23]. A multimodal broker realized via the SDO from [18] enables data exchange between the GFMI controllers for reliable higher-level coordination. Measurement data for the control are acquired from internal sensors within the parallel units, while data for the evaluation are recorded using a DEWETRON DEWE3-RM16 measurement system at a sampling rate of 200 kHz. The model and unit parameters are summarised in Table I.

### B. Parallel GFMI with GFLI as VAR Injector

In this scenario the Q–V behaviour of the two parallel GFMI units forming an island microgrid is identified. The PI

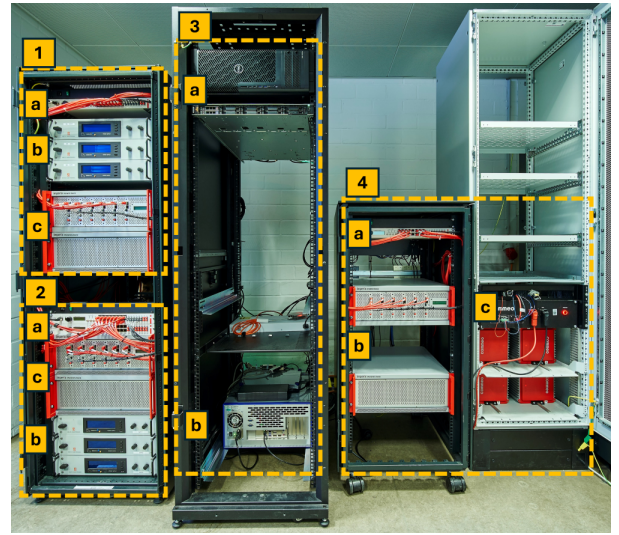

Fig. 3. Experimental setup consisting of four blocks: (1–2) GFMI$_1$ and GFMI$_2$, each including (a) controller, (b) three DC sources and (c) DC/DC and DC/AC stages; (3) communication unit with (a) SDO and (b) measurement unit; (4) GFLI comprising (a) controller, (b) DC/DC and DC/AC stages and (c) battery system.

TABLE I
CONTROL AND HARDWARE PARAMETERS

| Symbol | Value | Symbol | Value |
| --- | --- | --- | --- |
| $k_{\text{pvd}}, k_{\text{ivd}}$ | 0.187, 0.5 | $k_{\text{pvq}}, k_{\text{ivq}}$ | 0.52, 1.16 |
| $k_{\text{pid}}, k_{\text{iid}}$ | 6.25, 55 | $k_{\text{piq}}, k_{\text{iiq}}$ | 1, 10 |
| $k_{\text{drp}}, k_{\text{drq}}$ | 2e-4, 1e-3 | $\omega^*, v_{\text{m}}^*$ | $2\pi 50\,\text{Hz}, 230\sqrt{2}\,\text{V}$ |
| $k_{\text{p}\omega}, k_{\text{i}\omega}$ | 0.8, 0.8 | $k_{\text{pv}}, k_{\text{iv}}$ | 0.02, 0.3 |
| $k_{\text{fb}}, k_{\text{iQ}}$ | 0.7, 0.003 | $\rho_1 = \rho_2$ | 0.5 |
| $L_{\text{g}} = L_{\text{l}}$ | 2.2 mH | $C_{\text{g}}, f_{\text{s}}$ | $10\,\mu\text{F}, 20\,\text{kHz}$ |
| $v_{\text{dc}}^*$ | 700 V | $v_{\text{b}}, v_{\text{s}}$ | 400, 300 V |

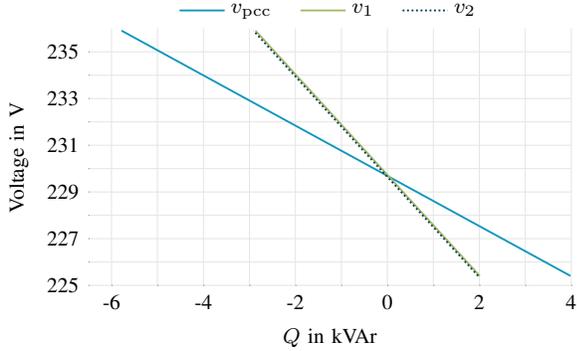

Fig. 4. Q–V characteristics of the baseline droop control case.

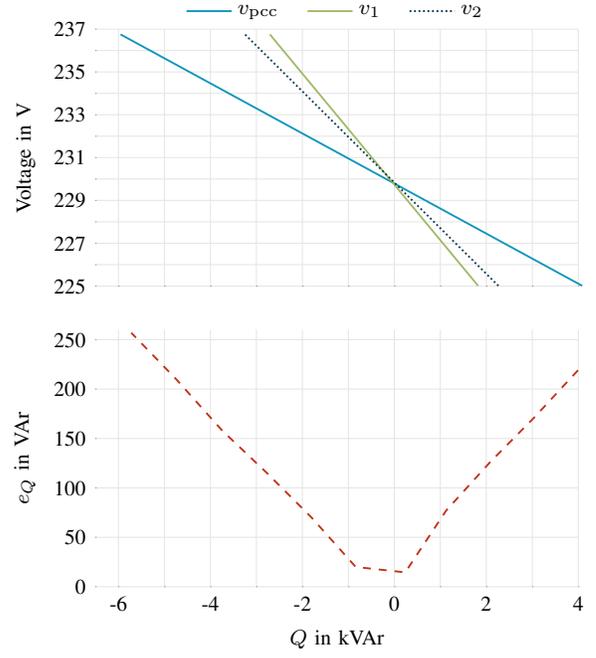

Fig. 5. Q–V characteristics and reactive power sharing error in the impedance mismatch case.

parameters, line impedances and output filters of both systems are identical. To assess their behaviour, the AC load power is set to 2 kW, while sweeping the GFLI reactive power setpoint $Q^*$ from $-5\,\text{kVAr}$ to $+5\,\text{kVAr}$ at a fixed active power setpoint of $P^* = 0\,\text{kW}$. Since the connection of the GFLI introduces a bias of approximately $-1\,\text{kVAr}$, the total reactive power contribution $Q_{\text{tot}}$ is effectively swept from $-6$ to $+4\,\text{kVAr}$. Figure 4 maps the measured fundamental PCC voltage to the total ($Q_{\text{tot}}$) and individual ($Q_1, Q_2$) measured reactive power supplied by the GFMIs. The results validate a linear Q–V characteristic. Furthermore, the individual responses of $Q_1$ and $Q_2$ match closely over the sweep range, with only a slight deviation of approximately $50\,\text{VAr}$, which can be attributed to minor hardware asymmetries in the experimental setup. It is important to note that the deviation of $v_{\text{pcc}}$ from the nominal value of 230 V is attributed to droop-induced voltage adjustment in addition to voltage drops across line inductances. Since this experiment provides a baseline for complete symmetric operation, the next experiment introduces a virtual impedance mismatch to demonstrate the reactive power sharing error in droop-only operation.

### C. Reactive Power Sharing under Impedance Mismatch

To introduce a line inductance mismatch, a virtual inductance $L_{\text{v}} = 1\,\text{mH}$ is added to GFMI$_2$. Hence, the droop equation is extended by the terms $\omega L_{\text{v}} i_{\text{q}}$ and $-\omega L_{\text{v}} i_{\text{d}}$ for $v_{\text{d}}^*$ and $v_{\text{q}}^*$, respectively. This modification alters the effective output impedance, which comprises the physical and virtual components, to demonstrate the resulting reactive power sharing error

$$e_Q = \frac{1}{2}|Q_1 - Q_2|. \tag{11}$$

Figure 5 shows the Q-V characteristics for the mismatched impedance scenario together with the corresponding reactive power sharing error $e_Q$. The virtual inductance changes the droop characteristic of GFMI$_2$ resulting in an increased reactive power contribution $Q_2$ compared to $Q_1$, which increases with higher inductive load. The reactive power sharing error reaches up to $0.2\,\text{kVAr}$ and $0.25\,\text{kVAr}$ at $4\,\text{kVAr}$ and $-6\,\text{kVAr}$ inductive load, respectively. The deviations motivate the introduction of additional reactive power sharing correction, which is employed in what follows.

### D. Distributed Reactive Power Sharing Control

In this scenario, the experimental setup is identical to the previous one, with the addition of the correction (10). The reactive power sharing correction can be manually activated simultaneously on both GFMIs. Deactivating it resets the integrator and replaces its output with a rate-limited reference that ramps the offset to zero to reduce additional transients when restoring baseline droop operation. To evaluate the error mitigation during dynamic operation, a reactive power transient is introduced. Hence, the reactive current reference $i_{\text{q}}^*$ of the GFLI is increased from 2 A to 10 A, resulting in a $4\,\text{kVAr}$ step.

Figure 6 shows the reactive power contribution of each GFMI as well as the corresponding mean reference value $(Q_1 + Q_2)/2$ before, during and after the transient event.

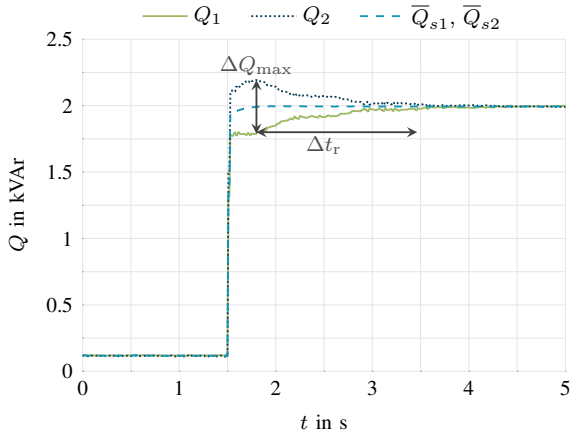

Fig. 6. Reactive power contribution during transient event.

TABLE II
MAXIMUM DEVIATION AND RECOVERY TIME FOR DIFFERENT INTEGRATORS

| $k_{iQ}$ | $\Delta Q_{max}$ | $\Delta t_r$ | Behaviour |
| --- | --- | --- | --- |
| 0.001 | 422.51 VAr | 6.76 s | Overdamped |
| 0.003 | 414.20 VAr | 1.74 s | Overdamped |
| 0.009 | 355.02 VAr | 1.48 s | Underdamped |
| 0.030 | 351.58 VAr | 2.92 s | Oscillating |
| 0.060 | – | – | Unstable |

Immediately after the event, the GFMIs dynamically redistribute their reactive power contribution with $Q_2$ temporarily exceeding and $Q_1$ remaining below the mean reference. The sharing deviation $\Delta Q = |Q_2 - Q_1|$ reaches its maximum value of 414.20 VAr during the redistribution phase. The restoration time $\Delta t_r$ is defined as the time required for $\Delta Q$ to decrease below 20 VAr for longer than one communication sampling period of the SDO (200 ms). In Fig. 6 we have $\Delta t_r = 1.74$ s with no remaining oscillations, showing the error mitigation capabilities of the correction term.

### E. Sensitivity Analysis of the Integral Gain

To evaluate the sensitivity of the integral gain, the transient experiment described in IV-D is repeated for different $k_{iQ}$. The transient event is initiated by raising the reactive power injection by the GFLI from zero to a step value of 4 kVAr. In total, six gains are investigated. The results are summarised in Table II. Increasing $k_{iQ}$ from 0.001 to 0.003 significantly reduces recovery time $\Delta t_r$ from 6.76 s to 1.74 s, while slightly decreasing the maximum sharing deviation $\Delta Q_{max}$. Further increasing the gain to 0.009 results in a faster response. However, the response becomes underdamped resulting in oscillations during the restoration phase. For $k_{iQ} = 0.03$, the increased control speed leads to a reduced $\Delta Q_{max}$, but increased oscillating behaviour increases the recovery time to 2.92 s. Finally, for $k_{iQ} = 0.06$ the system becomes unstable. Overall a gain $k_{iQ} = 0.003$ appears to provide the best compromise between convergence speed and damped behaviour for the reactive power sharing correction.

TABLE III
STABILITY CLASSIFICATION UNDER VARYING INTEGRAL GAIN AND COMMUNICATION DELAY

| | Delay in ms | | | | | |
| --- | --- | --- | --- | --- | --- | --- |
| $k_{iQ}$ | 50 | 60 | 70 | 80 | 90 | 100 |
| 0.001 | ● | ● | ● | ● | ● | ◐ |
| 0.003 | ● | ● | ● | ● | ● | ◐ |
| 0.009 | ● | ● | ● | ● | ● | ● |

● Stable ◐ Degraded ● Failed

### F. Robustness to Communication Delays and Packet Loss

In practical implementations of distributed control schemes, communication plays a crucial role. To evaluate the robustness of the reactive power sharing scheme to delays, the transient experiment (0 kVAr to 4 kVAr) is repeated with artificially introduced communication delays through the SDO. The implemented delays range from 50 ms to 100 ms in 10 ms steps and are evaluated for the gains $k_{iQ} \in \{0.001, 0.003, 0.009\}$. The communication rate is fixed at 5 Hz, corresponding to a sampling time of $T_s = 200$ ms, and the packet loss is set to 3 % for all experiments. The results are summarised in Table III. For delays up to 90 ms, reactive power sharing is maintained for all investigated gains, although smaller values of $k_{iQ}$ result in slower convergence. For a 100 ms delay, which corresponds to half the sampling time, the gain strongly affects the system response. For $k_{iQ} = 0.001$, the system remains stable, but exhibits slow convergence behaviour. For $k_{iQ} = 0.003$ the system oscillates and settles at a slightly reduced voltage level. Lastly for $k_{iQ} = 0.009$ the system converges to a poor steady-state with lasting sharing error and high voltage deviation. The fixed packet loss of 3 % does not noticeably affect the system behaviour, since in the case of occasional packet drops the controller holds the previous value. Consequently, the overall system performance is primarily dominated by the communication delay.

## V. CONCLUSION

A critical challenge in parallel operation of GFMIs is the accurate and robust sharing of reactive power among participating units. In this work, a distributed integral correction term for reactive power sharing is introduced into a system comprising two GFMIs, an AC load and a GFLI. The robustness of the proposed approach against communication-induced stressors and impedance mismatch is systematically investigated. The system behaviour under reactive power transients injected by the GFLI was evaluated for different integral gains and communication delays ranging from 50 ms to 100 ms under a fixed packet loss rate of 3 %. The results indicate a stability boundary between 90 ms and 100 ms communication delay. This bound decreases as the integral gain grows. These findings contribute to the ongoing discussion on grid-forming readiness, where grid-forming battery systems in distribution networks are the key contributors to resilience and standalone operation [24].

Future work will focus on the stochastic effects of increased packet loss and on extending the proposed approach toward a holistic control scheme by combining it with the operation control layer described in [25].